\documentstyle[preprint,eqsecnum,aps,floats,epsf]{revtex}
\begin{document}
\tighten
%
\draft
\date{\today}
\preprint{NSF-ITP-96-76,\ UCSBTH-94-27,\ DAMTP/96-37,\ LAUR-96-1117}
\title{Signature of the Simplicial Supermetric}

\author{James B. Hartle\thanks{Permanent address: Department of Physics, 
                               University of California, 
			       Santa Barbara, CA 93106,
                               hartle@itp.ucsb.edu}}
\address{Theoretical Division (T-6, MS B288), Los Alamos National 
Laboratory, Los Alamos, NM 87545}
\address{Santa Fe Institute, 1399 Hyde Park Rd, Santa Fe, NM 87501}
\address{Department of Physics \& Astronomy, University of New Mexico, 
Albuquerque, NM 87131}
\vskip .13 in
\author{Warner A. Miller\thanks{wam@lanl.gov}}
\address{Theoretical Division (T-6, MS B288), Los Alamos 
National Laboratory, Los Alamos, NM 87545}
\author{Ruth M. Williams\thanks{rmw7@damtp.cam.ac.uk}}
\address{DAMTP, Silver Street, Cambridge CB3 9EW, UK}
\address{Girton College, Cambridge CB3 0JG, UK}

\maketitle

\begin{abstract}
We investigate the signature of the Lund-Regge metric on spaces of
simplicial three-geometries which are important in some formulations
of quantum gravity.  Tetrahedra can be joined together to make a
three-dimensional piecewise linear manifold.  A metric on this
manifold is specified by assigning a flat metric to the interior of
the tetrahedra and values to their squared edge-lengths. The subset of
the space of squared edge-lengths obeying triangle and analogous
inequalities is simplicial configuration space.  We derive the
Lund-Regge metric on simplicial configuration space and show how it
provides the shortest distance between simplicial three-geometries
among all choices of gauge inside the simplices for defining this
metric (Regge gauge freedom).  We show analytically that there is
always at least one physical timelike direction in simplicial configuration
space and provide a lower bound on the number of spacelike directions.
We show that in the neighborhood of points in this space corresponding
to flat metrics there are spacelike directions corresponding to gauge
freedom in assigning the edge-lengths.  We evaluate the signature
numerically for the simplicial configuration spaces based on some
simple triangulations of the three-sphere $(S^3)$ and three-torus
$(T^3)$.  For the surface of a four-simplex triangulation of $S^3$ we
find one timelike direction and all the rest spacelike over all of the
simplicial configuration space.  For the triangulation of $T^3$ around
flat space we find degeneracies in the simplicial supermetric as well
as a few gauge modes corresponding to a positive eigenvalue. Moreover,
we have determined that some of the negative eigenvalues are physical,
i.e. the corresponding eigenvectors are not generators of
diffeomorphisms. We compare our results with the known properties of
continuum superspace.
\end{abstract}


\narrowtext
\setcounter{footnote}{0}
\section[]{Introduction}
\label{sec:I}

The superspace of three-geometries on a fixed manifold \cite{Whe68}
plays an important role in several formulations of quantum gravity.
In Dirac quantization \cite{DeW67}, wave functions on superspace
represent states.  In generalized quantum frameworks \cite{Harup},
sets of wave functions on superspace define initial and final
conditions for quantum cosmology. The geometry of superspace is
therefore of interest and has received considerable
attention.\footnote{For some representative earlier articles see
\cite{DeW67,Fis70,DeW70a}. For more recent articles that also review
the current situation see \cite{FH90,Giu95}.} The notion of distance
that defines this geometry is induced from the DeWitt supermetric on
the larger space of three-{\it metrics} \cite{DeW67,DeW70a}. While the
properties of the supermetric on the space of metrics 
are explicit, the properties of the
induced metric\footnote{The induced metric on the superspace of three
metrics might also be called the DeWitt metric since it was first
explored by DeWitt \cite{DeW70a}. However, to avoid confusion we
reserve the term ``DeWitt metric'' for the metric on the
space of three-metrics.} on the space of three-{\it geometries} are
only partially understood \cite{FH90,Giu95}. For example, the
signature of the metric on superspace, which is of special interest
for defining spacelike surfaces in superspace, is known only in
certain regions of this infinite dimensional space.  In this paper we
explore the signature of the metric on simplicial approximations to
superspace generated by the methods of the Regge calculus
\cite{Reg61}.

Tetrahedra (three-simplices) can be joined together to make a
three-dimensional, piecewise linear manifold.  A metric on this
manifold may be specified by assigning a flat metric to the interior
of the simplices and values to their $n_1$ squared edge-lengths.  Not
every value of the squared edge-lengths is consistent with a
Riemannian metric (signature +++) on the simplicial manifold.  Rather,
the squared edge-lengths must be positive, satisfy the triangle
inequalities, and the analogous inequalities for tetrahedra.  The
region of an ${\bf R}^{n_1}$ whose axes are the squared edge-lengths
$t^i$, $i=1, \cdots, n_1$, where these inequalities are satisfied
is a space of simplicial configurations we call {\em simplicial
configuration space}.\footnote{This is the
``truncated'' superspace of \cite{MTW}}
The DeWitt supermetric induces a metric on
simplicial configuration space. Lund and Regge \cite{LRup} have given
a simple expression for this metric 
and its properties have been explored by Piran and Williams
\cite{PW86} and Friedman and Jack \cite{FJ86}.  In this paper we
explore the signature of the Lund-Regge metric for several
simplicial manifolds by a combination of analytical and numerical
techniques.  In contrast to the continuum problem, we are able to
explore the signature over the whole of the finite dimensional
simplicial configuration spaces.

In Section II we review the construction of the metric on the
superspace of continuum three-geometries and summarize the known
information on its signature.  In Section III we show how the
Lund-Regge simplicial metric is induced from the continuum metric
and analytically derive a
number of results limiting its signature.  Section IV explores the
signature numerically for a number of elementary, closed, simplicial
manifolds.  We study first the surface of a four-simplex.  We find that
throughout its 10-dimensional configuration space that among 
a basis of orthogonal vectors there is one timelike
direction and 9 spacelike ones.  We next study the Lund-Regge metric of a
three-torus at various lattice resolutions (ranging from a
189-dimensional to a 1764-dimensional simplicial configuration space)
in the neighborhood of the single point representing a flat metric.
We find that the Lund-Regge metric can be degenerate, change signature,
and have more than one physical time-like directions.  
We conclude with a comparison with known continuum results.

\section[]{Continuum Superspace}
\label{sec:II}

In this section we shall briefly review some of the known properties of
the metric on the superspace of continuum geometries on a fixed
three-manifold $M$. We do this to highlight the main features that we
must address when analyzing the corresponding metric on the superspace 
of simplicial geometries.  A more detailed account of the continuum situation
can be readily found \cite{Whe68,DeW67,Fis70,DeW70a,FH90,Giu95}.

Geometries on $M$ can be represented by three-metrics $h_{ab}(x)$,
although, of course, different metrics 
describe the same geometry when related by a diffeomorphism. 
We denote the space of
three-metrics on $M$ by ${\cal M}(M)$. A point in ${\cal M}$ is a
particular metric $h_{ab}(x)$ and we may consider the tangent space of
vectors at a point. Infinitesimal displacements $\delta h_{ab}(x)$ from
one three-metric to another are particular examples of vectors.  We
denote such vectors generally by $k_{ab}(x)$, $k^\prime_{ab}(x)$, etc. A
natural class of metrics on ${\cal M}(M)$ emerges from the structure of
the constraints of general relativity. Explicitly they are given by
\begin{equation}
(k^\prime,k) = \int\nolimits_M d^3 x N(x) \bar G^{abcd} (x) k^\prime_{ab}(x)
k_{cd}(x)
\label{twoone}
\end{equation}
where $\bar G^{abcd}(x)$, called the inverse DeWitt supermetric, is
given by
\begin{equation}
\bar G^{abcd}(x) = 
\frac{1}{2} h^{\frac{1}{2}}(x) \left[h^{ac} (x) h^{bd} (x) +
h^{ad} (x) h^{bc} (x) - 2 h^{ab} (x) h^{cd}(x)\right]
\label{twotwo}
\end{equation}
and $N(x)$ is an essentially arbitrary but non-vanishing function called
the lapse.  Different metrics result from different choices of $N(x)$. In
the following we shall confine ourselves to the simplest choice,
$N(x)=1$.

The DeWitt supermetric (\ref{twotwo}) at a point $x$ defines a metric on the
six-dimensional space of three-metric components at $x$.  This metric
has signature $(-,+,+,+,+,+)$ \cite{DeW67}. The signature of the metric
(\ref{twoone}) on ${\cal M}$ therefore has an infinite number of
negative signs and an infinite number of positive signs --- roughly one
negative sign and five positive signs for each point in $M$.

The space of interest, however, is not the space of
three-metrics ${\cal M}(M)$ but rather the superspace of
three-geometries Riem$(M)$ whose ``points'' consist of classes of
diffeomorphically equivalent metrics, $h_{ab}(x)$.  A metric on Riem$(M)$
can be induced from the metric on ${\cal M}(M)$, (\ref{twoone}), by
choosing a particular perturbation in the metric $\delta h_{ab}(x)$ to
represent the infinitesimal displacement between two nearby
three-geometries.  However, a $\delta h_{ab}$ is not fixed uniquely by
the pair of nearby geometries.  Rather, as is well known, there is an
arbitrariness in $\delta h_{ab}(x)$ corresponding to the arbitrariness
in how the points in the two geometries are identified. That
arbitrariness means that, for any vector $\xi^a(x)$, 
the ``gauge-transformed'' perturbation
\begin{equation}
\delta h^\prime_{ab}(x) = \delta h_{ab}(x) + D_{(a}\xi_{b)}(x)\ ,
\label{twothree}
\end{equation}
represents the same displacement in superspace as $\delta h_{ab}(x)$
does,  where  $D_a$ is the derivative in $M$.

The metric (\ref{twoone}) on ${\cal M}(M)$ is not invariant under gauge
transformations of the form (\ref{twothree}) even with $N=1$.  Thus we
may distinguish ``{\it vertical}'' directions in ${\cal M}(M)$ which are pure
gauge
\begin{equation}
k^{\rm vertical}_{ab} (x) = D_{(a}\xi_{b)}(x)
\label{twofour}
\end{equation}
and ``{\it horizontal}'' directions which are orthogonal to {\it all} of these
in the metric (\ref{twoone}).

Since the  metric (\ref{twoone}) is not invariant under gauge
transformations, there are different notions of distance between points
in superspace depending on what $\delta h_{ab}(x)$ is used to represent
displacements between them.  The conventional choice
\cite{DeW70a} for
defining a geometry on superspace has been to choose the {\it minimum}
of such distances between points.  That is the same as saying that
distance is measured in ``horizontal'' directions in superspace.
Equivalently, one could say that a gauge for representing displacements
has been fixed. It is the gauge specified by the three conditions
\begin{equation}
D^b \left(k_{ab} - h_{ab} k^c_c\right) = 0\ .
\label{twofive}
\end{equation}

The signature of the metric defined by the above construction is an
obvious first question concerning the geometry of superspace.  The
infinite dimensionality of superspace, however, makes this a non-trivial
question to answer.  The known results have been lucidly explained by
Friedman and Higuchi \cite{FH90} and Giulini \cite{Giu95} and we briefly
summarize some of them here: 
\begin{itemize}
	\item At any point in superspace there is always at least one negative 
	direction represented by constant conformal displacements of the form
		\begin{equation}
			k_{ab} = \delta\Omega^2  h_{ab}(x)\ .
		\label{twosix}
		\end{equation}
	Evidently (\ref{twosix}) satisfies (\ref{twofive}) so that it is
	horizontal, and explicit computation from (\ref{twoone}) shows $(k,
	k) \leq 0$; 

	\item If $M$ is the sphere $S^3$, then for a
	neighborhood of the round metric on $S^3$, the signature has one
	negative sign corresponding to (\ref{twosix}) and all other orthogonal 
	directions are positive; 

	\item Every $M$ admits geometries with negative Ricci curvature
	(all eigenvalues strictly negative).  In the open region of
	superspace defined by negative Ricci curvature geometries the signature
	has an {\it infinite} number of negative signs and an {\it infinite}
        number of positive signs.  On the sphere, these results already
	show that there must be points in superspace where the metric is
	degenerate.
\end{itemize}

The above results are limited, covering only a small part of the
totality of superspace.  In the following we shall show that
more complete results can be obtained in simplicial configuration space.

\section[]{The Lund-Regge Metric}
\label{sec:III}

\subsection{Definition}
\label{sec:A}

In this Section we derive the form of the Lund-Regge metric on
simplicial configuration space together with some analytic results on its
signature.  We consider a fixed closed simplicial three-manifold $M$
consisting of $n_3$ tetrahedra (three-simplices) joined together so that
each neighborhood of a point in $M$ is homeomorphic to a region of ${\bf
R}^3$. The resulting collections of $n_k$ $k$-simplices, (vertices,
edges, triangles, and tetrahedra for $k=0,1,2,3$, respectively) we
denote by $\Sigma_k$.  A simplicial geometry is fixed by an assignment
of values to the squared edge-lengths of $M,\ t^m, m=1, \cdots, n_1$ and
a flat Riemannian geometry to the interior of each tetrahedron consistent
with those values.  The assignment of squared edge-lengths is not
arbitrary.  The squared edge-lengths are positive and constrained by the
triangle inequalities and their analogs for the tetrahedra.  Specifically
if $V^2_k(\sigma)$ is the squared measure (length, area, volume) of
$k$-simplex $\sigma$ expressed as a function of the $t^m$, we must have
\begin{equation}
V^2_k(\sigma)\geq 0\ , \quad k=1,2,3
\label{threeone}
\end{equation}
for all $\sigma\in \Sigma_k$.  The space of three-geometries on $M$ is
therefore the subset of the space of $n_1$ squared edge-lengths $t^m$
in which (\ref{threeone}) is satisfied.  We call this {\it simplicial
configuration space} and denote it by ${\cal T}(M)$. A point in ${\cal
T}(M)$ is a geometry on $M$; the $\{t^m\}$ are coordinates locating
points in  ${\cal T}(M)$.

Distinct points in ${\cal T}(M)$ correspond to different assignments
of edge-lengths to the simplicial manifold $M$.  In general distinct
points correspond to distinct three-geometries and,  in this respect,
${\cal T}(M)$ is like a superspace of three-geometries.  However, this
is not always the case.  Displacements of the vertices of a flat
geometry in a flat embedding space result in a new assignment of the
edge-lengths that corresponds to the same flat geometry.  These
variations in edge-lengths that preserve geometry are the simplicial
analogs of diffeomorphisms \cite{Har85,Mor92}.  Further, for large
triangulations where the local geometry is near to flat we expect
there to be approximate simplicial diffeomorphisms --- small changes
in the edge-lengths which approximately preserve the geometry
\cite{Har85,LNN86,RW84}.  Thus, the continuum limit of ${\cal T}(M)$
is not the superspace of three-geometries but the space of
three-metrics.  It is for these reasons that we have used the term
simplicial configuration space rather than simplicial superspace.

We now define a metric on ${\cal T}(M)$ that gives the distance
between points separated by infinitesimal displacements $\delta t^m$
according to
\begin{equation}
\delta S^2 = G_{mn} (t^\ell) \delta  t^m \delta  t^n\ .
\label{threetwo}
\end{equation}
Such a metric can be induced from the DeWitt metric on the space
${\cal M}(M)$ of continuum three-metrics on $M$ in the following way:

Every simplicial geometry can be represented in ${\cal M}(M)$ by a
metric which is piecewise flat in the tetrahedra and, indeed, there are
many different metrics representing the same geometry. Every
displacement $\delta t^m$ between two nearly three-geometries can be
represented by a perturbation $\delta h_{ab}(x)$ of the metric in ${\cal
M}(M)$.  The DeWitt metric (\ref{twoone}) which gives the notion of
distance between nearby metrics in ${\cal M}(M)$ can therefore be used
to induce a notion of distance in ${\cal T}(M)$ through the relation
\begin{equation}
G_{mn} (t^\ell) \delta t^m \delta t^n = \int\nolimits_M d^3 x\, N(x)
\,\bar G^{abcd}(x) \delta h_{ab}(x) \delta h_{cd}(x)\ .
\label{threethree}
\end{equation}
On the right hand side $\bar G^{abcd}(x)$ is (\ref{twotwo}) evaluated at
a piecewise flat metric representing the simplicial geometry and $\delta
h_{ab}(x)$ is a perturbation in the metric representing the change in
that geometry corresponding to the displacement $\delta t^m$.

However, as the discussion of Section II should make clear, many
different metrics on ${\cal T}(M)$ can be induced by the
identification (\ref{threethree}).  First, there is the choice of
$N(x)$.  We choose $N(x)=1$.  Second, since the right hand side of
(\ref{threethree}) is not gauge invariant we must fix a gauge for the
perturbations $\delta h_{ab}$ to determine a metric $G_{mn} (t^\ell)$.
There are two parts to this. First, to evaluate the integral on the right
hand side of (\ref{threethree}) we must at least fix the gauge {\it
inside} each tetrahedron.  We shall refer to this as the {\sl Regge
gauge} freedom.  It is important to emphasize, however, that any
choice for the Regge gauge does not completely fix the total gauge
freedom available.  As discussed above, there still may be variations
of the lengths of the edges which preserve the geometry -- simplicial
diffeomorphisms --- and correspondingly $\cal T(M)$ can still have
both vertical and horizontal directions. Therefore, secondly, this 
gauge freedom must also be fixed. 
  
A natural choice for the Regge gauge from the point of view of
simplicial geometry is to require that the $\delta h_{ab}$ are {\it
constant} inside each tetrahedron,
\begin{equation}
D_c \delta h_{ab} (x) = 0\ , \quad {\rm inside\ each}\quad
 \tau\in \Sigma_3\ ,
\label{threefour}
\end{equation} 
but possibly varying from one tetrahedron to the next.  The conditions
(\ref{threefour}) are, of course, more numerous than the three
diffeomorphism conditions permitted at each point, but as we already
mentioned, these two gauges are distinct.  Thus, (\ref{threefour}) is
not the Regge calculus counterpart of (\ref{twofive}). Nevertheless,
this choice of Regge-gauge (\ref{threefour}) has a beautiful property
which we shall discuss below in Subsection B.

Assuming (\ref{threefour}), the right hand side of (\ref{threethree})
may be evaluated explicitly.  Although not gauge invariant, the
right-hand-side of (\ref{threethree}) {\em is} coordinate invariant.  We
may therefore conveniently use coordinates in which the metric
coefficients $h_{ab}(x)$ satisfying (\ref{threefour}) are
constant in each tetrahedron. Then,
using (\ref{twotwo}),
\begin{equation}
G_{mn}(t^\ell)\delta t^m\delta t^n =
 \sum\limits_{\tau\in\Sigma_3}
V(\tau)\left\{\delta h_{ab} (\tau) \delta h^{ab}(\tau) - \left[\delta
h^a_a(\tau)\right]^2\right\}
\label{threefive}
\end{equation}
where $V(\tau)$ is the volume of tetrahedron $\tau$ and we have written
$h_{ab}(\tau), \delta h_{ab}(\tau)$, etc for the constant values of
these tensors inside $\tau$.

To proceed further we need explicit expressions for $h_{ab}(\tau)$ in
terms of $t^\ell$, and for $\delta h_{ab}(\tau)$ in terms of $t^\ell$ and
$\delta t^\ell$.  One way of making an explicit identification is to
pick a particular vertex in each tetrahedron $(0)$ and consider the
vectors ${\bf e}_a(\tau), a= 1,2,3$ proceeding from this vertex to the other
three vertices $(1, 2, 3)$ along the edges of the tetrahedron.  The metric
$h_{ab}(\tau)$ in the basis defined by these vectors is
\begin{eqnarray}
\label{threesix}
h_{ab}(\tau) &=& {\bf e}_a(\tau)\cdot {\bf e}_b(\tau)\nonumber \\
&=& \frac{1}{2} \left(t_{0a}+t_{0b}-t_{ab}\right)  
\end{eqnarray}
where $t_{AB}$ is the squared edge-length between vertices
$A$ and $B$.  Eq (\ref{threesix}) gives an explicit expression
for the metric in each tetrahedron in terms of its squared edge-lengths in a
basis adapted to its edges.  The perturbation of (\ref{threesix}),
\begin{equation}
\delta h_{ab} (\tau) = \frac{1}{2} \left(\delta t_{0a} + \delta t_{ab} - \delta
t_{0b}\right) \ , 
\label{threeseven}
\end{equation}
gives an explicit expression for the perturbation in $h_{ab}(\tau)$
induced by changes in the squared edge-lengths.  (In general
(\ref{threeseven}) changes discontinuously from tetrahedron to
tetrahedron.)  Eq (\ref{threeseven}) is an explicit realization of the
gauge condition (\ref{threefour}). Only trivial linear transformations
of the form (\ref{twothree})  inside the tetrahedra preserve (\ref{threefour}),
and there are none of these that preserve the simplicial structure in
the sense that $\xi^a(x)$ vanishes on the boundary of the tetrahedra.
In this sense (\ref{threeseven}) fixes the Regge gauge for the
perturbations. 

An explicit expression for the $G_{mn}(t^\ell)$ defined by
(\ref{threethree}), (\ref{threefive}), and (\ref{threeseven}) may be
obtained by studying the expression\footnote{See, {\it e.g.}
 \cite{Har85} for a
derivation.} for the squared volume of tetrahedron $\tau$,
\begin{equation}
V^2(\tau) = \frac{1}{(3!)^2} {\rm det} [h_{ab}(\tau)]\ .
\label{threeeight}
\end{equation}
Consider a perturbation $\delta t^m$ in the squared edge-lengths.  The
left hand side of (\ref{threeeight}) may be expanded (dropping the label
$\tau$) as
\begin{equation}
V^2 (t^\ell + \delta t^\ell)=
 V^2(t^\ell) + \frac{\partial
V^2(t^\ell)}{\partial t^m} \delta t^m + \frac{1}{2}\ \frac{\partial^2 V^2
(t^\ell)}{\partial t^m \partial t^n}\ \delta t^m \delta t^n + \cdots\ .
\label{threenine}
\end{equation}
The right hand side may be expanded using the identity
\begin{equation}
{\rm det}\ A = \exp[Tr \log(A)] 
\label{threeten}
\end{equation}
as
\begin{eqnarray}
{\rm det} \left(h_{ab} + \delta h_{ab}\right) 
&=& {\rm det} (h_{ab}) {\rm det} \left(\delta^a_b + \delta h^a_b\right) \nonumber \\
&=& {\rm det} (h_{ab}) \left\{1 + \delta h^a_a 
                                + \frac{1}{2}\left[\delta h_{ab} \delta h^{ab} 
                                - \left(\delta h^a_a\right)^2\right] 
                                +\cdots \right\}\ .
\label{threeeleven}
\end{eqnarray}
Equating (\ref{threenine}) and (\ref{threeeleven}) gives, at first order,
the identity
\begin{equation}
\delta h^a_a(\tau) = \frac{1}{V^2}\ \frac{\partial V^2(\tau)}{\partial
t^m}\ \delta t^m\ ,
\label{threetwelve}
\end{equation}
and at second order gives a relation which leads through (\ref{threefive}) to
the following elegant expression for the metric $G_{mn} (t^\ell)$:
\begin{equation}
G_{mn} (t^\ell) = -\sum\limits_{\tau\in\Sigma_3} \frac{1}{V(\tau)} 
\ \frac{\partial^2 V^2(\tau)}{\partial t^m\partial t^n}\ .
\label{threethirteen}
\end{equation}
This is the Lund-Regge metric \cite{LRup} on simplicial
configuration space ${\cal T}(M)$. It is an explicit function of the squared 
 edge-lengths $t^m$ through (\ref{threeeight}) and (\ref{threesix}). The metric
may be reexpressed in a number of other ways of which a useful example
is 
\begin{equation}
G_{mn} (t^\ell) = -2 \left[\frac{\partial^2 V_{\rm TOT}}{\partial t^m
\partial t^n} + \sum\limits_{\tau\in\Sigma_3} \frac{1}{V(\tau)} \frac{\partial
V(\tau)}{\partial t^m}\ \frac{\partial V(\tau)}{\partial t^n}\right]
\label{threefourteen}
\end{equation}
where $V_{\rm TOT}$ is the total volume of $M$.

The metric (\ref{threefourteen}) becomes singular at the boundary of
$\cal T(M)$ where $V(\tau)$ vanishes for one or more tetrahedra.
However, locally, since $V^2$ is a third order polynomial in the $t$'s, 
$G_{mn} \sim  (t^\ell - t^\ell_b)^{-1/2}$ where $t^\ell_b$ is a point
on the boundary.  A generic boundary point is therefore a finite distance
from any other point in ${\cal T(M)}$ as measured by the metric $G_{mn}$. 

\subsection{Comparison with Nearby Regge Gauge Choices}
\label{sec:B}

The identification of points in a perturbed and unperturbed geometry
is ambiguous up to a displacement $\xi^a(x)$ in the point in the
perturbed geometry identified with $x^a$ in the unperturbed geometry.
As a consequence any  two perturbations $\delta h_{ab}(x)$ which
differ by a gauge transformation (\ref{twothree}) represent the same
displacement in the space ${\cal T}(M)$ of three-geometries.  In the
continuum case of Section II, we followed DeWitt \cite{DeW70a} and
fixed this ambiguity by minimizing the right-hand-side of
(\ref{threethree}) over all possible gauge transformations $\xi^a(x)$
so that distance between three-geometries was measured along
``horizontal'' directions in ${\cal M}(M)$.  In the  previous subsection 
we fixed
the ambiguity in the comparison of the continuum space of piecewise metrics
to simplicial lattices by requiring perturbations to be constant over
tetrahedra [{\it cf.}  (\ref{threefour})].  We can now show that the
distance defined in this way is a local minimum with respect to other
Regge gauge choices that preserve the simplicial structure, in a sense to be
made precise below.

Consider the first variation of the right-hand-side of
(\ref{threethree}) with $N(x)=1$ that is produced by an infinitesimal
gauge transformation
$\xi^a(x)$. This is
\begin{equation}
\int_M d^3 x \bar G^{abcd} (x) \delta h_{ab}(x) D_{(c}\xi_{d)}(x)\ .
\label{threefifteen}
\end{equation}
Integrating by parts and making use of the symmetry of $\bar G^{abcd}$, this
first variation can be written 
\begin{equation}
-\sum\limits_{\tau\in\Sigma_3} \int_\tau d^3 x \bar G^{abcd} (x)
D_c \delta h_{ab} (x) \xi_d(x) +  
\sum\limits_{\sigma\in\Sigma_2} \int_\sigma d\Sigma\left[\hskip-.01
in\Bigl|n_c
\bar G^{abcd} (x) \delta h_{ab} (x)\Bigr |\hskip-.01 in\right] 
\xi_d (x)\ .
\label{threesixteen}
\end{equation}
In this expression, the first term is a sum of volume integrals over
the individual tetrahedra in $M$.  The second term is an integral over
triangles where, for a particular triangle, $n^a$ is a unit outward pointing
normal and
$\left[\hskip-.01 in\Bigl|\ \ \Bigr |\hskip-.01 in\right]$ denotes the
discontinuity across the triangle.  Such a term must be included since
we do not necessarily assume that the non-gauge invariant argument of
(\ref{threefifteen}) is continuous from tetrahedron to tetrahedron.  The
conditions (\ref{threefour}) make the first term vanish.  The second
vanishes when $\xi^a(x)$ vanishes on the boundary of every
tetrahedron.  That means that the distance defined by the Lund-Regge
metric is an extremum among all re-identifications of points in the
{\it interiors} of the tetrahedra between the perturbed and
unperturbed geometries.  It does not appear to necessarily be an
extremum with respect to re-identifying points in the interior of
triangles or edges.  The Lund-Regge metric therefore provides the
shortest distance between simplicial three-metrics among all 
choices of Regge gauge which vanish on the triangles.
However it is  not exactly ``horizontal'' in the sense of the continuum
because of the possibility of simplicial diffeomorphisms.
We shall see explicit consequences of this below.

\subsection{Analytic Results on the Signature}
\label{sec:C}

We are interested in the signature of $G_{mn}$ on ${\cal T}(M)$.  We shall
calculate the signature numerically for some simple $M$ in
Section \ref{sec:IV}, but here we give a few analytic results which
characterize it incompletely.

\subsubsection{The Timelike Conformal Direction }
\label{sec:a}

The conformal perturbation defined by
\begin{equation}
\delta t^m = \delta{\Omega^2} \ t^m
\label{threeseventeen}
\end{equation}
is always timelike.  This can be seen directly from (\ref{threefive})
by noting that (\ref{threeseven}) and (\ref{threesix}) imply
\begin{equation}
\delta h_{ab}(\tau) = \delta {\Omega^2} \ h_{ab}(\tau)\ .
\label{threeeighteen}
\end{equation}
However, it can also be verified directly from (\ref{threethirteen})
using the fact that $V^2$ is a homogeneous polynomial of degree three
in the $t^m$.  Then it follows easily from Euler's theorem that
\begin{equation}
G_{mn} (t^\ell) t^m t^n = - 6V_{\rm TOT} (t^\ell) < 0\ .
\label{threenineteen}
\end{equation}

The timelike conformal direction is not an eigenvector of $G_{mn}$
because
\begin{equation}
G_{mn}t^n=-4 \partial V_{TOT}/\partial  t^m\  .
\label{conformal}
\end{equation}
We do not expect $\partial V_{TOT}/\partial t^m $ to be proportional
to $t^m$ except for symmetric assignments of the edge lengths on 
highly symmetric triangulations. 

The same relation shows that the conformal direction is orthogonal
to any gauge direction $\delta t^n$ because
\begin{equation}
G_{mn}t^m\delta t^n=-4 (\partial V_{TOT}/\partial  t^n) \delta t^n =
-4\delta V_{TOT}\ . 
\label{conformal1 }
\end{equation}
This vanishes for any change in edge lengths which does not change the
geometry. 

\subsubsection{At Least $n_1-n_3$ Spacelike Directions}
\label{sec:b}

Eqs (\ref{threefive}) and (\ref{threetwelve})  can be combined to show
that
\begin{eqnarray}
\widetilde G_{mn} \delta t^m \delta t^n &\equiv& \left[G_{mn} + 4
\sum\limits_{\tau\in\Sigma_3} \, \frac{1}{V(\tau)}\ \frac{\partial V(\tau)}{\partial
t^m}\ \frac{\partial V(\tau)}{\partial t^n}\right] \delta
t^m\delta t^n \nonumber \\
&=&\sum\limits_{\tau\in\Sigma_3} \, V(\tau)\left[\delta h_{ab} (\tau)
\delta h^{ab}(\tau)\right]\geq 0
\label{threetwenty}
\end{eqnarray}
Thus
\begin{equation}
G_{mn} = \widetilde G_{mn} - 4\sum\limits_{\tau\in\Sigma_3}\ \frac{1}{V(\tau)}
\ \frac{\partial V(\tau)}{\partial t^m}
\ \frac{\partial V(\tau)}{\partial t^n}
\label{threetwentyone}
\end{equation}
where $\widetilde G_{mn}$ is positive.  Some displacements $\delta t^m$
will leave the volumes of all the tetrahedra unchanged:
\begin{equation}
\frac{\partial V(\tau)}{\partial t^m}\ \delta t^m = 0\quad , \quad \tau
\in \Sigma_3.
\label{threetwentytwo}
\end{equation}
These directions are clearly spacelike from (\ref{threetwentyone}).
Since (\ref{threetwentytwo}) is $n_3$ conditions on $n_1$ displacements
 $\delta t^m$
we expect at least $n_1-n_3$ independent spacelike directions.

\subsubsection{Signature of Diffeomorphism Modes}
\label{sec:c}

In general any change $\delta t^m$ in the squared edge-lengths of $M$
changes the three-geometry. A flat simplicial three-geometry is an
exception. Locally a flat simplicial geometry may be embedded in
Euclidean ${\bf R}^3$ with the vertices at positions ${\bf x}_A, A=1,
\cdots, n_0$. Displacements of these locations result in new and
different edge-lengths, but the flat geometry remains unchanged.  Such
changes in the edge-lengths $\delta t^m$ are called {\sl gauge directions} in
simplicial configuration space.  Each vertex may be displaced in three
directions making a total of $3n_0$ gauge directions.  We shall now
investigate whether these directions are timelike, spacelike, or null.

We evaluate $\delta S^2$ defined by (\ref{threetwo}) and
(\ref{threefourteen}) for displacements $\delta{\bf x}_A$ in the
locations of the vertices.  If an edge connects vertices $A$ and $B$,
its length is
\begin{equation}
t^{AB} = \left({\bf x}_A - {\bf x}_B\right)^2 \equiv \left({\bf
x}_{AB}\right)^2
\label{threetwentythree}
\end{equation}
and the change in length $\delta t^{AB}$ from a variation in
position $\delta {\bf
x}_A$ follows immediately.  The total volume is unchanged by any
variation in position of the ${\bf x}_A$, which means that
\begin{equation}
\frac{\partial V_{TOT}}{\partial  x^i_A} = 0 \qquad , \qquad
\frac{\partial^2 V_{TOT}}{\partial x^i_A \partial x^j_B} = 0 
\label{threetwentyfour}
\end{equation}
and so on.  These derivatives are related to those with respect to the
edge-lengths by the chain rule.
%
Thus (\ref{threetwentyfour}) does not imply that $\partial^2
V_{TOT}/\partial t^{AC}\partial t^{BC}$ is zero, but only that 
\begin{equation}
\frac{\partial^2 V_{TOT}}{\partial t^m \partial t^n} \delta t^m \delta
t^n = -\frac{\partial V_{TOT}}{\partial t^{AB}}
\ \frac{\partial^2t^{AB}}{\partial x^i_A \partial x^j_B} \delta x^i_A
\delta x^j_B
\label{threetwentyseven}
\end{equation}
where $\delta t^m = (\partial t^m/\partial x^i_A )\delta
x^i_A$, with summation over both $A$ and $i$. Inserting 
(\ref{threetwentyseven}) and the chain rule relations into 
(\ref{threefourteen}), we obtain
\begin{eqnarray}
\lefteqn{\delta S^2 \equiv G_{mn} (t^{\ell}) \delta t^m \delta t^n =} 
\\ \nonumber
- 2 \Biggl[-\frac{\partial
V_{TOT}}{\partial t^n}\ \frac{\partial^2t^n}{\partial x^i_A \partial
x^j_B} 
&+ &  \sum\limits_{\tau\in\Sigma_3} \frac{1}{V(\tau)}
\ \frac{\partial V(\tau)}{\partial t^m}\ \frac{\partial V(\tau)}{\partial t^n}
\ \frac{\partial t^m}{\partial x^i_A}\ \frac{\partial t^n}{\partial
x^j_B}\Biggr] \delta x^i_A \delta x^j_B\ . 
\label{threetwentyeight}
\end{eqnarray}
To simplify this, we go from sums over edges to sums over the
corresponding vertices, $(C, D)$, with a factor of $1/2$ for each sum.
Then
\begin{equation}
\delta S^2 = \frac{\partial V_{TOT}}{\partial t^{CD}}
\ \frac{\partial^2t^{CD}}{\partial x^i_A \partial x^j_B}\ \delta x^i_A 
\delta x^j_B -
\frac{1}{2} \sum\limits_{\tau\in\Sigma_3} \frac{1}{V(\tau)} 
\left[\frac{\partial V(\tau)}{\partial t^{CD}}
\ \frac{\partial t^{CD}}{\partial x^i_A} \delta x^i_A\right]^2\ .
\label{threetwentynine}
\end{equation}
Using the explicit relations (\ref{threetwentythree}), we find
\begin{equation}
\delta S^2   =   2\left\{ \frac{\partial V_{TOT}}{\partial t_{CD}} \left( \delta 
{\bf x}_C - \delta {\bf x}_D\right)^2 
- \sum\limits_{\tau\in\Sigma_3}\ \frac{1}{V(\tau)} 
\left[\frac{\partial V(\tau)}{\partial t_{CD}}
{\bf x}_{CD} \cdot \left(\delta {\bf x}_C - \delta {\bf
x}_D\right) \right]^2\right\}.
\label{threethirty}
\end{equation}
The second term of (\ref{threethirty}) is negative definite, but
the first does not appear to have a definite sign, so that a general
statement on the character of gauge modes does not emerge. 
However, more information is available in the specific cases to 
be considered below. 

\section[]{Numerical Investigation of the Simplicial Supermetric}
\label{sec:IV}

The Lund-Regge metric can be evaluated numerically to give complete
information about its signature over the whole of the simplicial
configuration space which confirms the incomplete but more general
analytic results obtained above. We consider specifically as manifolds the 
three-sphere ($S^3$), and the three-torus ($T^3$).  For the simplest
triangulation of $S^3$ we investigate the signature over the whole of
its simplicial configuration space.  For several triangulations of
$T^3$ we investigate a limited region of their simplicial
configuration space near flat geometries.  Even though in both cases
the triangulations are rather course, a number of basic and
interesting features emerge.
 
The details of the triangulations in the two cases are described below
but the general method of calculation is as follows.  Initial
edge-lengths are assigned (consistent with the triangle and
tetrahedral equalities) and the supermetric calculated using (3.13).
The eigenvalues of the metric $G_{mn}$ are then calculated and the
numbers of positive, negative and zero values counted.  To explore
other regions of the simplicial configuration space, the edge-lengths
are repeatedly updated (in such a manner so as to ensure that the
squared-measures (see (3.1)) of all the triangles and tetrahedra are
positive) and the eigenvalues and hence the signature of the
supermetric are found.  In the case of the $T^3$ triangulations we
also calculate the deficit angles which give information on the curvature
of the simplicial geometry, and in addition, we explore the geometry
of the neighboring points in simplicial configuration space along each
eigenvector.

We now describe the details of the two numerical calculations for the two 
different manifolds. 

\subsection{The 3-Sphere}
\label{subsec:IVA} 
 
	The simplest triangulation of $S^3$ is the surface of a
four-simplex, which consists of five vertices, five tetrahedra, ten
triangles, and ten edges.  
Thus the  simplicial configuration
space is 10-dimensional (Figure 1).  Each point in this space
represents a particular assignment of lengths to each of the ten edges
of the 4-simplex.  To numerically explore all of this 10-dimensional
space would be foolish as the space contains redundant and ill-defined
regions. To avoid the ill-defined parts we need only to restrict
ourselves to those points in the configuration space that satisfy the
two and  three dimensional simplicial inequalities (Eq.~3.1).  However, 
to avoid  redundancy we must examine the invariance properties of the 
the eigenvalue spectrum of the Lund-Regge metric.

	Let us begin by examining an invariance of the Lund-Regge
eigenvalue spectrum under a global scale transformation. The
Lund-Regge metric scales as $G_{mn}\longrightarrow L^{-1}G_{mn}$ under
an overall rescaling of the edges $l_i\longrightarrow Ll_i$ where
$l_i=\sqrt{t^i}$.  The signature is scale invariant, so we may use
this invariance to impose one condition on the $l_i$.  We found it
most convenient numerically to fix $l_0 = 1$ as the longest edge.  The
ten-dimensional space has now collapsed to a nine-dimensional
subspace. As this is the only invariance we have identified, we then
further restrict our investigation to the points in this subspace
which satisfy the various simplicial inequalities. Specifically,
writing $\overline{AB}$ as the edge between vertex $A$
and vertex $B$:
\begin{itemize}
\item $l_0 \equiv \overline{34}=1$ and $l_0$ is the longest edge;

\item $l_3 \equiv \overline{03} \in (0,1]$;

\item $l_4 \equiv \overline{04} \in [(1-l_3),1]$;

\item $l_6 \equiv \overline{13} \in (0,1]$;

\item $l_7 \equiv \overline{14} \in [(1-l_6),1]$;

\item $l_1 \equiv \overline{01} \in (\tau_{-}, \tau_{+})$ where
      $\tau_\pm$ is obtained from the tetrahedral inequality applied to 
      edge $\overline{01}$ of tetrahedron $(0134)$,
\begin{eqnarray}
\tau_\pm^2 & = & \frac{1}{2} \left(-t_0+t_3+t_4+t_6-\frac{t_3t_6}{t_0}
+\frac{t_4t_6}{t_0} \right. \nonumber \\
& + &  t_7+\frac{t_3t_7}{t_0}-\frac{t_4t_7}{t_0} \nonumber \\
& & \pm \sqrt{(t_0^2-2t_0t_3+t_3^2-2t_0t_4-2t_3t_4+t_4^2)} \nonumber \\
& & \left. \sqrt{(t_0^2-2t_0t_6+t_6^2-2t_0t_7-2t_6t_7+t_7^2)}/{t_0} \right)
\label{fourone}
\end{eqnarray}

\item $l_8 \equiv \overline{23} \in (0,1]$;

\item $l_9 \equiv \overline{24} \in [(1-l_8),1]$;

\item $l_2 \equiv \overline{02} \in (\tau_{-}, \tau_{+})$, where
      $\tau_\pm$ is obtained from a tetrahedral inequality applied to 
      edge $\overline{02}$ of tetrahedron $\{0234\}$;

\item $l_5 \equiv \overline{12} \in (s_{-},s_{+})$, where $s_\pm$ is 
      obtained from the tetrahedral inequality as applied to the three
      tetrahedra ( $(1234)$, $(0123)$ and $(0124)$ ) sharing edge
      $\overline{12}$, with
\begin{eqnarray}
s_- & = & max\{ \tau_-^{(1234)}, \tau_-^{(0123)}, \tau_-^{(0124)} \} 
\nonumber \\
s_+ & = & min\{ \tau_+^{(1234)}, \tau_+^{(0123)}, \tau_+^{(0124)} \}.
\label{fourtwo}
\end{eqnarray}    
\end{itemize}
Considering only this region, we were able to sample the whole of the
configuration space.  Subdividing the unit interval into ten points
would ordinarily entail a calculation of the eigenvalues for $10^{10}$
points. However, by using the scaling law for the Regge-Lund metric
together with the various simplicial inequalities we only needed to
calculate the eigenvalues for $102160$ points. We observed exactly one
timelike direction and nine spacelike directions for each point in the
simplicial configuration space, even though the distortions of our
geometry from sphericity were occasionally substantial  --- up to 10 to 1
deviations in the squared edge-lengths from their symmetric values.
Recall that a conformal displacement is a timelike direction in
simplicial configuration space [{\it cf.}
(\ref{threenineteen})]. However this direction will coincide with the
timelike eigenvector of $G_{mn}$ only when all the edge lengths are
equal as follows from (\ref{conformal}).

There are clearly more than the $n_1-n_3=5$ spacelike
directions required by the general result of \ref{sec:b}.  We conclude
that in this case the signature of the Lund-Regge metric is
$(-,+,+,+,+,+,+,+,+)$ over the whole of simplicial configuration
space.

\begin{figure}
\centerline{\epsfxsize=5.0truein\epsfbox{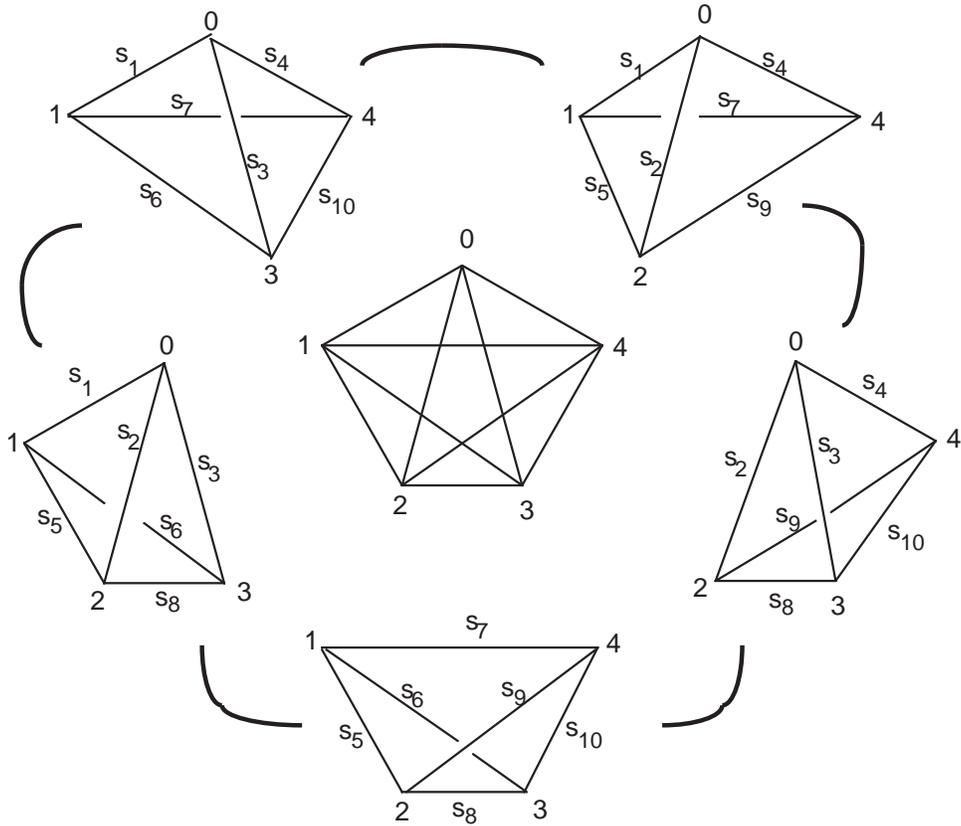}}
\label{Fig1}
\caption{\protect\small
{\em The boundary of a 4-simplex as a 10-dimensional simplicial
configuration space model for $S^3$.}  This figure shows the five
tetrahedra corresponding to the boundary of the central 4-simplex
(0,1,2,3,4) exploded off around its perimeter.  The 4-simplex has 5
vertices, 10 edges, 10 triangles, and 5 tetrahedra.  The topology of
the boundary consisting of those tetrahedra is that of a 3-sphere.
The specification of the 10 squared edge-lengths of the 4-simplex
completely fixes its geometry, and represents a single point in the
10-dimensional simplicial configuration space.  Here we analyze the geometry of
this space using the Regge-Lund metric and show that there is one, and
only one, timelike direction.} 
\end{figure}

More detailed information about the Lund-Regge metric beyond the
signature is contained in the eigenvalues themselves.  Predictions of
their degeneracies arise from the symmetry group of the triangulation.

For the boundary of the 4-simplex the symmetry group is the
permutation group on the five vertices, $S_5$.  If all the edge
lengths are assigned symmetrically --- all equal edge-lengths in the
present case --- then the eigenvalues may be classified according to
the irreducible representations of $S_5$ and their degeneracies are
given by the dimensions of those representations.  This is because a
permutation of the vertices can be viewed as a matrix in the
10-dimensional space of edges which interchanges the edges in
accordance with the permutation of the vertices.  The Regge-Lund
metric $G_{mn}$ may be viewed similarly and commutes with all the
elements of $S_5$ for a symmetric assignment of edges.  The matrices
representing the elements of $S_5$ give a 10-dimensional reducible
representation of it, which can be decomposed into irreducible
representations by standard methods described in \cite{Har85c}.  The
result is that the reducible representation decomposes as $1+4+5=10$,
where the factors in this sum are dimensions of the irreducible
representations, which we expect to be the multiplicities of the
corresponding eigenvalues of $G_{mn}$ at a symmetric assignment of
edges.

The results of numerical calculations of the eigenvalues are illustrated in
Figure~2 for a slice of simplicial configuration space.  When
all the edges are equal to 1 we found one eigenvalue of $-1/2\sqrt{2}$,
four eigenvalues equal to $1/3\sqrt{2}$, and five of $5/6\sqrt{2}$. As
expected, these degeneracies were broken when we departed from the
spherical geometry (Figure~2). Nevertheless, even with aspect ratios on
the order of $10:1$ we always found a single timelike direction in
this simplicial superspace.

\begin{figure}
\centerline{\epsfxsize=5.0truein\epsfbox{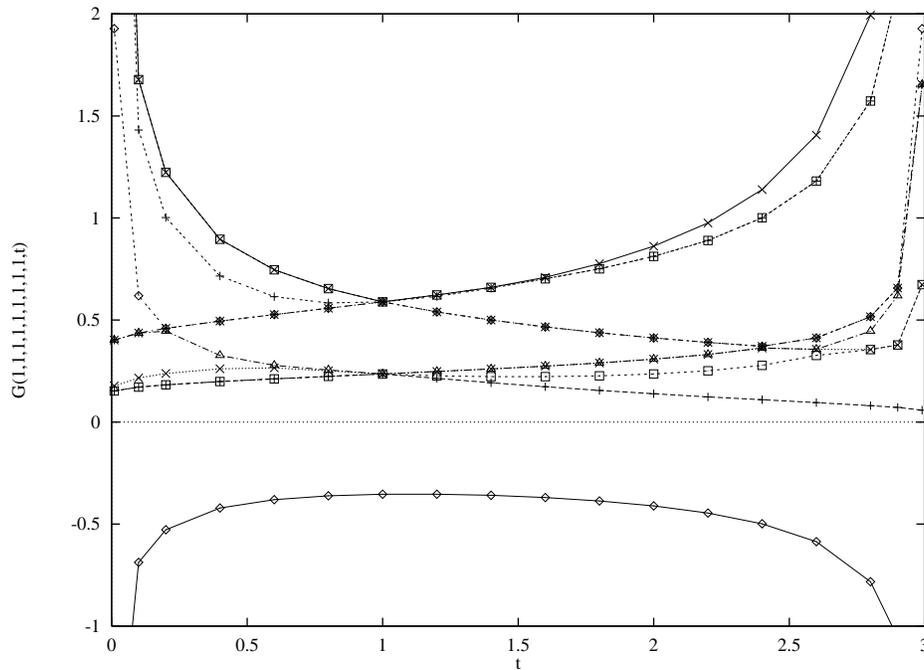}}
\label{Fig2}
\caption{\protect\small
{\em The eigenvalue spectrum along a 1-dimensional cut through the
10-dimensional simplicial configuration space.}  We plot the 10
eigenvalues (not necessarily distinct) of the Lund-Regge metric
$G_{mn}$ with nine of the ten edges set to unity and the remaining
squared edge-length ($t$) varies from 0 to 3 (a range obtained
directly from the tetrahedral inequality).  When $t=1$ we have maximal
symmetry and we have three distinct eigenvalues for the ten
eigenvectors as expected with degeneracies 1, 4, and 5 as predicted.
As we move away from this point one can see that the degeneracies are
for the most part broken.  However, there remains a twofold degeneracy
in the 5 arising from the remaining symmetries in the assignments of
the edge-lengths.  }
\end{figure}

\section{The 3-Torus}
\label{subsec:IVB}

In this section we analyze the Lund-Regge metric in the neighborhood
of two different flat geometries on a common class of triangulations of the
three-torus, $T^3$. We investigate the metric on triangulations
of varying refinement in this class.
We illustrate degeneracy of the metric and identify a few gauge modes
(vertical directions) in flat space which
correspond to positive eigenvalues.
 
The class of triangulations of $T^3$ is constructed as follows:  A lattice of
cubes, with $n_x$, $n_y$ and $n_z$ cubes in the $x$, $y$ and
$z$-directions, is given the topology of a 3-torus by identifying
opposite faces in each of the three directions.  Each cube is divided
into six tetrahedra, by drawing in face diagonals and a body diagonal
(for details, see R\v ocek and Williams\cite{RW81}).  The number of
vertices is then $n_0=n_x n_y n_z$ and there are $7n_0$ edges, $12n_0$
triangles, and $6n_0$ tetrahedra.  

The geometry is flat when the  squared edge lengths of the 
sides, face diagonals, and body
diagonals take values of
1, 2, and 3 times the lattice scale respectively.     
We refer to this geometry as the {\em
right-tetrahedron lattice}.

The flat 3-torus can also be tessellated by isosceles tetrahedra, each
face of which is an isosceles triangle with a squared base edge of
$1$, the other two squared edge-lengths being $3/4$. We refer to this
as the {\em isosceles-tetrahedron lattice}.
One can obtain this lattice from the right-tetrahedron lattice by the
following construction on each cube. Compress the
cube along its main diagonal in a symmetrical way, keeping all the ``coordinate
edges" at length $1$, until the main diagonal is also of length $1$. The face
diagonals will then have squared edge-lengths of $4/3$. An overall rescaling by a
factor of $\sqrt 3/2$ then converts the lattice of these deformed cubes to the
isosceles-tetrahedron lattice.
 
These two lattices correspond to distinct points in the $7n_0$ dimensional
simplicial configuration space of a triangulation of $T^3$. 
They are examples of inequivalent flat geometries on $T^3$.
Even though they can be obtained from each other by the deformation
procedure described, they are associated with distinct flat structures. For
example, consider the geodesic structure. For the right-tetrahedron lattice,
there are three orthogonal geodesics of extremal length at any point, parallel
to the coordinate edges of the cubes of which the lattice is constructed, and
corresponding to the different meridians of the torus. On the other hand,
for the isosceles-tetrahedron lattice, the three extremal geodesics at any
point will not be orthogonal to each other; they will actually be parallel to
the ``coordinate edges" which in the deformation process have moved into
positions at angles of ${\rm arccos}(-1/3)$ to each other. (It is perhaps easier to
visualize the analogous situation in two dimensions where the geodesics will
be at angles of ${\rm arccos}(-1/2)$ to each other). If the metric structures on the two
lattices we consider were diffeomorphically equivalent, the diffeomorphism
would preserve geometric quantities, like the angles between the
extremal-length geodesics, and this is clearly not the case. 
We shall see how
this inequivalence between the flat tori manifests itself in the simplicial
supermetric.

We first turn to a detailed examination of the the isosceles-tetrahedron
lattice and, in particular, the eigenvalues of the Lund-Regge metric.
Unlike the relatively simple $S^3$ model described above where we used
Mathematica to calculate the eigenvalue spectrum of the matrix
$G_{mn}$, here we developed a C program utilizing a Householder method
to determine the the eigenvalue spectrum and corresponding
eigenvectors.  In addition, we calculated the deficit angle
(integrated curvature) associated to each edge. These deficit angles
were used to identify diffeomorphism and conformal directions as well
as to corroborate the analytic results for the continuum described in
the Introduction.  We performed various runs ranging from an
isosceles-tetrahedron lattice with $3\times 3\times 3$ vertices and
189 edges, up to a lattice with $6\times 6\times 7$ vertices and 1764
edges. The points (simplicial 3-geometries) in such high dimensional
simplicial configuration spaces cannot be systematically canvassed as
we did in the 4-simplex model. For this reason we chose to search the
neighborhood of flat space in two ways. First we explored the region
around flat space by making random variations (up to 20 percent) in
the squared edge-lengths of the isosceles-tetrahedron lattice, and
secondly we perturbed the edge-lengths a short distance along selected
flat-space eigenvectors.

Movement along the eigenvectors was performed in the following way. We
start with a set of squared edge-lengths corresponding to zero
curvature.  All the $7n_0$ deficit angles are zero. We calculate the
$7n_0$ eigenvectors, $v=\{v_j,\ j=1,2,\ldots,7n_0\}$ together with the
corresponding eigenvalues $\lambda_j$ and then adjust the squared
edge-lengths along one of the $v_j$ as specified in:
\begin{equation}
t^i_{new} = t^i_{flat} + \epsilon v^j,\hspace{.25in} \forall i\in 
\{1,2,\ldots, 7n_0\}.
\label{foureight}
\end{equation}
Here we choose $\epsilon\ll 1$.  We then calculate the deficit angles
for this new point $t_{new}$ and then repeat this procedure for each
$v_j$ in turn. In this way we can in principle identify which (if any)
of the eigenvectors correspond to gauge directions. 
We observe the following:
\begin{itemize}
  \item Eigenvectors corresponding to  eigenvalues $\lambda=1/2$
	appear to be diffeomorphisms to order $\epsilon^3$ in the 
        sense that the deficit angles are of  order $\epsilon^3$.
  \item For an $n_0=n\times n\times n$ lattice there are $6n-4$
	eigenvectors corresponding to $\lambda=1/2$.
  \item There are $n_0+2$ eigenvectors corresponding to $\lambda=1$.
\end{itemize}
A graphical representation of this movement along the eigenvectors away from 
the flat isosceles-tetrahedron lattice point is illustrated in Figure~3.

\begin{figure}
\centerline{\epsfxsize=5.0truein\epsfbox{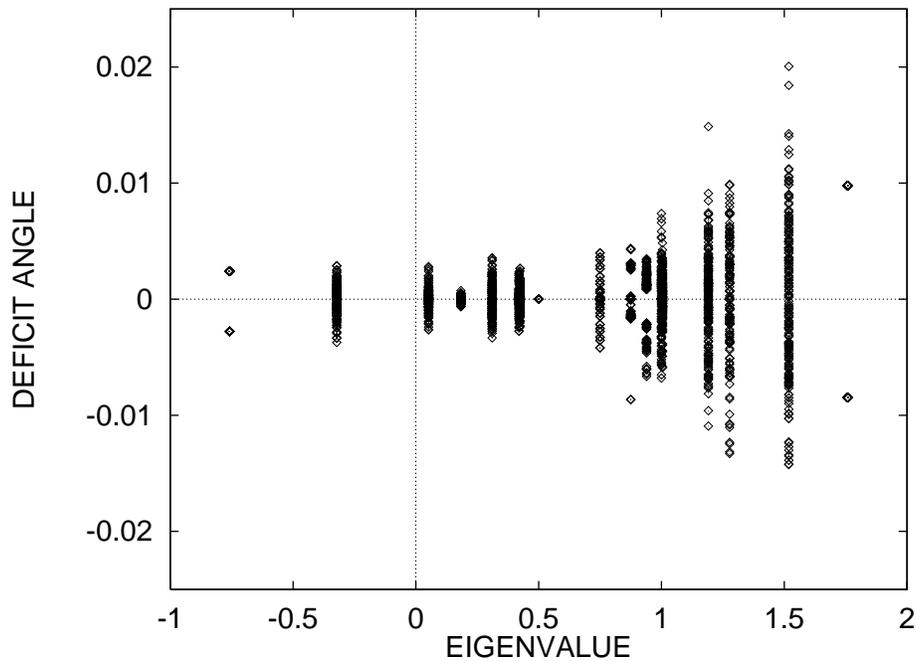}}
\label{Fig3}
\caption{\protect\small
{\em The generation of curvature by motion in simplicial configuration
space from a flat-space point to a neighboring point along one of the
flat-space eigenvectors.}  Here we analyze the curvature of the
3-geometries of the $3\times 3\times 3$ $T^3$ lattice in the
neighborhood of the flat-space isosceles tetrahedral lattice point.
The simplicial configuration space is 189 dimensional, and there are
189 deficit angles (integrated curvature) used as indicators of
curvature change.  This plot represents the deficit angle spectrum
associated to motion along each of the eigenvectors associated to each
of the eigenvalues $\lambda_k$  Eq.~(\protect\ref{foureight}). Here we chose
$\epsilon=0.01$. In the plot we notice that all the eigenvectors
corresponding to eigenvalue $\lambda=1/2$ correspond to vertical or
diffeomorphism directions.  }
\end{figure}

Although the eigenvalues corresponding to those gauge directions we have
identified is positive, we have not been able to establish all gauge
directions are spacelike. 
For the right-tetrahedron
lattice we have simplified the expression (\ref{threethirty}) obtained
for diffeomorphism modes as follows. The derivatives of $V(\tau)$ with
respect to the diagonal edges all vanish, so that only the edges of
the cubic lattice need be included in the $(C, D)$ sum.  For lattice
spacing of 1, these derivatives are all $1/12$, giving $\partial
V_{TOT}/\partial t^{CD} \equiv 1/2$, since each such edge is shared by
6 tetrahedra.  Thus for this lattice we have
\begin{equation} 
\delta S^2 = 
 \sum_{(C,D)} \left(\delta {\bf x}_C -  \delta {\bf
x}_D\right)^2 - \frac{1}{12}\ \sum_{\tau\in\Sigma_3} 
\left[\sum\limits_{(C,D)\in
\tau} {\bf x}_{CD} \cdot \left(\delta {\bf x}_C - \delta {\bf
x}_D\right) \right]^2 
\label{fourthree}
\end{equation}
where $(C,D)$ implies that $C$ and $D$ are connected by an edge.  
Alternatively, in
terms of summation over edges, $m$, the expression is
\begin{equation}
\delta S^2 = 2\left[\sum_m \delta {\bf x}^2_m  -\frac{1}{6}
\sum_{\tau\in\Sigma_3} \left(\sum_{m\in\tau} {\bf x}_m
\cdot \delta {\bf x}_m\right)^2\right]
\label{fourfour}
\end{equation}

Even though (\ref{fourfour}) is relatively simple, with definite
signs for each term, we have not managed to prove a general result about
the overall sign. We suspect that it is positive and in the following
special cases have found this to
be true:
\begin{enumerate}
\item If just one vertex moves through $\delta {\bf r}$
\begin{equation}
\delta S^2=  8 \delta {\bf r}^2  
\label{fourfive}
\end{equation}
\item If only the edges in one coordinate direction of the lattice
change,
\begin{equation}
\delta S^2 \ge  \frac{1}{2} \sum \delta {\bf x}^2_{CD} 
\label{foursix}
\end{equation}
\item If all the $\delta {\bf x}_{CD}$'s have the same magnitude
$\delta L$ (but not the same direction)
\begin{equation}
\delta S^2 > (\delta L)^2\ \left(n_1 - \frac{3}{4}\ n_3\right) > 0  
\label{fourseven}
\end{equation}
since $n_1 > n_3$
\end{enumerate}

If it could be proved that $\delta S^2$ is always positive, it would
still remain to calculate the number of independent directions in the
space of edge-lengths, in order to see how many of the positive
eigenvalues do indeed correspond to vertical gauge modes.

We next turn briefly to more general aspects of the 
right-tetrahedron lattice. 
As mentioned earlier, although the geometry of the right-tetrahedron is flat 
like the isosceles-tetrahedron lattice, it is not diffeomorphic to it.
It is therefore no surprise that when we calculate the eigenvalues of
the Lund-Regge metric we find both their values and degeneracies to differ
from the isosceles-tetrahedron case.  
Even if the lattices are rescaled so that their total volumes are
equal, the eigenvalue spectra (which scale by the inverse length) are still not
the same. 

The right-tetrahedron case presents allows a particularly easy analysis
of how the degeneracies of the eigenvalue spectrum are connected with
the symmetry group of a lattice. To find the symmetry group of the
right-tetrahedron lattice,
consider the symmetries when one point, the origin say, is
fixed. These are a ``parity" transformation (when ${\bf r}$ is mapped
to $-{\bf r}$), represented by $Z_2$, and a permutation of the three
coordinate directions, represented by $S_3$. The full group is
obtained by combining this $Z_2 \times S_3$ subgroup with the subgroup
of translations (mod 3) in the three coordinate directions. Thus the
symmetry group of the $T^3$ lattice with $3\times 3\times 3$ vertices
is the semi-direct product of the elementary Abelian normal subgroup
of order $3^3$ by the subgroup $Z_2 \times S_3$.

The action of the group on the vertices induces a permutation of the edges.
This 189-dimensional permutation representation of the edges decomposes as
\begin{eqnarray}
3\times 1_1 + 2\times 2_1 &+& 3\times 2_2 + 2\times 4 +  5
\times (6_1 + 6_2 + 6_3 + 6_4) +  2  \times (6_5 + 6_6 + 6_7 + 6_8)
\label{fournine}
\end{eqnarray} 
where {\it e.g.} $6_3$ is the third irreducible representation of
dimension $6$. When this is compared with the multiplicities found for
the eigenvalues of the supermetric for the flat-space decomposition
with right-angled tetrahedra, it can be seen that the numbers agree
precisely provided that the two multiplicities of $8$ are interpreted
as $6 + 2$, and the two multiplicities of $3$ are regarded as $2 +
1$. We have no explanation for this unexpected degeneracy, although it
has been observed before (for example for several triangulations of
$CP^2$ \cite{Har85c,HPup}) and we suspect that there is a deep group
theoretical reason for it. A detailed investigation of the eigenvalues
found for the isosceles-tetrahedron lattice would almost certainly
reveal similar accidental degeneracies, for example for the
multiplicity of 29 found numerically for the eigenvalue $\lambda=1$.

Finally we look at common properties of the isosceles-tetrahedron
lattice and right tetrahedron lattice, in particular, the
the signature of the Lund-Regge metric, which is
the main point of these calculations.  For both the isosceles- and
right-tetrahedron lattices with $3\times 3\times 3$ vertices, and
therefore 189 edges, the supermetric (in a flat-space configuration,
and in a neighborhood of flat space) has 176 positive eigenvalues and
13 negative ones (Fig.~4 illustrates these results for the
isosceles-tetrahedron lattice, together with the other 32 runs we
made).  These are consistent with our analytical results and can be
interpreted as follows.
\begin{itemize}
	\item There are rather more than the required $n_1-n_3=27$
              spacelike directions.  
	\item The negative eigenvalues
              include the conformal mode.  
	\item We have shown that the eigenvalues corresponding to
              diffeomorphisms may sometimes be positive, and for the
              isosceles-tetrahedron lattice have identified each of
              the eigenvectors corresponding to the $\lambda=1/2$
              eigenvalue as a generator of a diffeomorphism, or
              vertical direction.
	\item None of the eigenvectors corresponding to the $13$ negative 
	      eigenvalues are generators of diffeomorphisms. This 
              indicates that there are $13$ horizontal directions corresponding to
              negative eigenvalues.
\end{itemize}
Furthermore, we consistently observe signature change in the
supermetric as we depart from the flat space point.  We also observed
a few null eigenvalues for the right-tetrahedron lattice for flat
space at various resolutions, including $4\times 4\times4$.  We are
presumably seeing finite analogs of the infinite number of horizontal
(i.e.  non-gauge) directions predicted by Giulini \cite{Giu95} for
regions of superspace in the neighborhood of a flat metric, where
there is an open region with negative Ricci curvature.  This
illustrates the important fact that there are still timelike
directions in the simplicial supermetric, beyond the conformal mode,
which are always present.

\vskip .26 in
\begin{figure}
\centerline{\epsfxsize=5.0truein\epsfbox{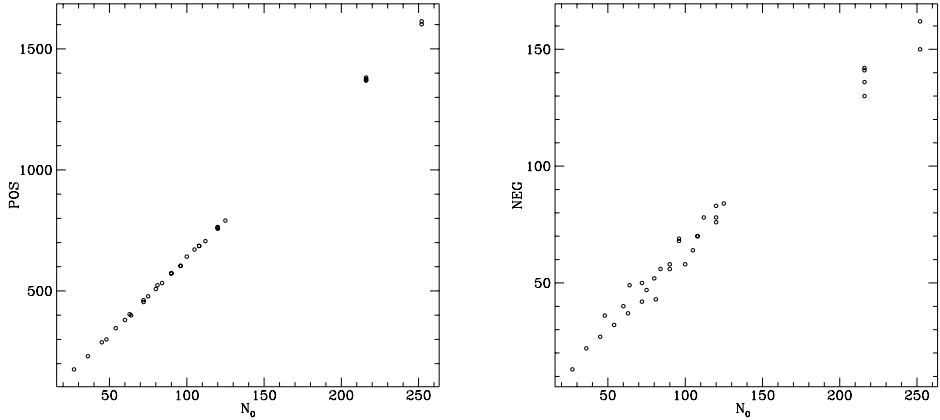}}
\label{Fig4}
\caption{\protect\small
{\em An originally flat 3-torus isosceles-tetrahedron lattice with
additional 10\% random fluctuations induced on the squared
edge-lengths.}  These two graphs show the number of positive ($POS$)
and negative ($NEG$) eigenvalues as a function of the number of
vertices ($N_0$). Here we considered 33 different resolution $T^3$
lattices ranging from a $3\times 3\times 3$ lattice with $N_0=27$
vertices to a 6x6x7 lattice with $N_0=252$ vertices.  The ratio of
negative to positive eigenvalues ranges from $\sim 0.074$ to $\sim
0.123$.} 
\end{figure}

\section{Geometric Structure of Superspace: Results and Future Directions}
\label{sec:V}

In this paper we used the simplicial supermetric of Lund and Regge as
a tool for analyzing the geometry of simplicial configuration space,
specifically the signature of the Lund-Regge metric.  One way of
summarizing our results is to compare them with the known results for
the metric on continuum superspace described in the Introduction:

The conformal direction (\ref{threeseventeen}) is timelike as we showed
analytically in (\ref{threenineteen}). This coincides with the result 
for the continuum conformal displacements.

The simplest simplicial manifold with the topology of the three-sphere
($S^3$) is the surface of a four-simplex.  Here, we showed that, in
the 10-dimensional simplicial configuration space, among a set of
orthogonal directions there is always a single timelike direction and
nine spacelike ones, even for regions of simplicial configuration
space corresponding to geometries distorted from spherical symmetry
with aspect ratios exceeding $10:1$ in squared edge-length. 
For the continuum such a result
is known only in an arbitrarily small neighborhood of the round metric
(analogous to all equal edges).  However, we have no evidence that the
situation of a single timelike direction in an orthogonal set extends to more
refined triangulations of $S^3$ such as the 600-cell.  In particular, 
preliminary results indicate that there are 628 positive, 92 negative 
and no zero eigenvalues for the 600-cell model\cite{BHM96}.

By investigating a neighborhood of the flat geometry in various
triangulations of $T^3$, we exhibited exact simplicial diffeomorphisms
for exactly flat geometries, and approximate simplicial
diffeomorphisms for approximately flat assignments of the squared edge
lengths. We showed that that there was more than one orthogonal
timelike non-gauge direction at the flat geometry. We showed that the
Lund-Regge metric can become degenerate and change signature as one
moves away from exactly flat geometries --- a result that might have
been expected at least on large triangulations of $S^3$ from 
the combination of the continuum
results on the signature near a round metric and the different
signature on regions that correspond to negative curvature. Here,
signature change is exhibited explicitly for $T^3$. 

The principle advantage of casting the DeWitt supermetric into its
simplicial form is to reduce the continuum infinite dimensional
superspace to a finite dimensional simplicial configuration
space. This simplicial configuration space is to be contrasted with
``mini-'' or ``midi''-superspaces. Simplicial configuration space preserves
elements of {\it both} the physical degrees of freedom and the
diffeomorphisms.  In the continuum limit of increasingly large
triangulations we expect to recover the full content of both.

Our analysis provides motivation for further research. A potentially
fruitful line of investigation is to define approximate notions of
vertical and horizontal directions in simplicial configuration space
in such a way that they coincide with the exact vertical and
horizontal directions in the continuum limit.  We already know that
there are $3n_0$ approximate diffeomorphism degrees of freedom for a
simplicial 3-geometry -- a fact that has been demonstrated
analytically and illustrated numerically in Regge geometrodynamics via
the freedom of choice of a shift vector per vertex\cite{Mil86,KMW88}.

Once armed with such a theory of approximate simplicial
diffeomorphisms, it would be interesting to extend our $S^3$ analysis
to a simplicial model with arbitrarily large number of vertices.  In
this way we can be assured that the tessellation will have encoded in
it all of the true dynamic degrees of freedom as well as the full
diffeomorphism freedom.

\section{Acknowledgments}

We are indebted to Benjamin Bromley for his invaluable assistance with
the numerical implementation of the $T^3$ lattice and for numerous
discussions. We are grateful to Jan Saxl for his analysis of the
multiplicities for the flat $T^3$ tessellation. JBH thanks the LANL
Theoretical Division, the Santa Fe Institute, and the Physics
Department at the University of New Mexico for their hospitality while
this work was started.  His work was supported in part by NSF grants
PHY90-08502, PHY95-07065, and PHY-94-07194.  RMW and WAM acknowledge
support from a Los Alamos National Laboratory LDRD/IP grant. The work
of RMW was also supported in part by the UK Particle Physics
and Astronomy Research Council.

\end{document}